\documentclass[12pt,a4paper]{article}
\usepackage{graphicx}
\usepackage{authblk}

\usepackage{amsmath}
\usepackage{amsfonts}
\usepackage{amssymb}
\usepackage{graphicx}

\newcommand{\beq}{\begin{equation}}
\newcommand{\eeq}{\end{equation}}
\newcommand{\beqa}{\begin{eqnarray}}
\newcommand{\eeqa}{\end{eqnarray}}

\begin{document}

\title{Construction of the mutually orthogonal extraordinary supersquares}

\author[1]{Cristian Ghiu}
\author[2]{Iulia Ghiu\thanks{iulia.ghiu@g.unibuc.ro (corresponding author)}}

\affil[1]{{\small University Politehnica of Bucharest, Faculty of Applied Sciences,
Department of Mathematics II, Splaiul Independen\c tei 313, R-060042
Bucharest, Romania}}
\affil[2]{{\small Centre for Advanced Quantum Physics, Department of Physics,
University of Bucharest, PO Box MG-11, R-077125,
Bucharest-M\u{a}gurele, Romania}}

\maketitle

\begin{abstract}
Our purpose is to determine the complete set of mutually orthogonal squares of order $d$, which are not necessary Latin.
In this article, we introduce the concept of supersquare of order $d$, which is defined with the help of its generating subgroup in $\mathbb{F}_d\times \mathbb{F}_d$. We present a method of construction of the mutually orthogonal supersquares. Further, we investigate the orthogonality of extraordinary supersquares, a special family of squares, whose generating subgroups are extraordinary.  The extraordinary subgroups in $\mathbb{F}_d\times \mathbb{F}_d$ are of great importance in the field of quantum information processing, especially for the study of mutually unbiased bases. We determine the most general complete sets of mutually orthogonal extraordinary supersquares of order 4, which consist in the so-called Type I and Type II.
The well-known case of $d-1$ mutually orthogonal Latin squares is only a special case, namely Type I.
\end{abstract}



\section{Introduction}

The study of Latin squares is of great interest in different branches of mathematics, such as combinatorics, statistical experimental design, error correcting codes, graph theory. Numerous results on the mutually orthogonal Latin squares (MOLS) have been published over decades: their construction \cite{Bose}, perfect Latin squares \cite{Kim}, a generalized equivalence between the MOLS and affine planes \cite{Mullen-1992}, partially orthogonal Latin squares \cite{Mullen-2004}, mutually orthogonal frequency squares \cite{Mullen-2006}, mutually orthogonal equitable Latin squares \cite{Asplund}, orthogonal arrays \cite{Mah}, just to enumerate a few references in this field.

The research topic of MOLS has recently attracted attention in the young field of quantum information theory: the mean king's problem \cite{Hayashi}, quantum error correction codes \cite{Aly}, mutually unbiased bases \cite{Rao}, \cite{Paterek-2009}, \cite{Paterek-2010}.

A possible way of constructing the set of mutually unbiased bases is by generating classes of unitary operators, whose eigenvectors represent these bases. In other words, one needs to determine $d$ + 1 classes of $d-1$ commuting operators. These special operators are called mutually unbiased operators \cite{Band}. Two operators commute if their associated elements $v_1=(x_1,y_1)$ and $v_2=(x_2,y_2)$ in $\mathbb{F}_d\times \mathbb{F}_d$ satisfy $tr(x_1y_2-x_2y_1)=0$ \cite{Klimov-2007}, \cite{Ghiu-2012}, \cite{Ghiu-2013}. The subgroup of $d$ elements in $\mathbb{F}_d\times \mathbb{F}_d$, whose any two elements satisfy the above condition is the so-called extraordinary subgroup. This subgroup is the key element for analyzing the problem of mutually unbiased bases since it generates the set of $d-1$ commuting operators.

Our paper is organized as follows. In Sec. 2 we present the definition of a square of order $d$, which is based on the partition of $\mathbb{F}_d\times \mathbb{F}_d$. Further, the concept of supersquare of order $d$ is introduced in Sec. 3 and the method of obtaining the complete set of mutually orthogonal supersquares is given. The extraordinary supersquares are defined in Sec. 4.1. Sec. 4.2 presents the construction of all the complete sets of mutually orthogonal extraordinary supersquares of order 4. These sets consist in the so-called Type I and Type II. Examples of Type I and II complete sets of mutually orthogonal extraordinary supersquares of order 4 are given in Sec. 4.3. The construction of the mutually extraordinary supersquares may have applications in the field of quantum information theory. Finally, we make some concluding remarks in Sec. 5.

\section{Preliminaries}

 A square of order $d$, denoted by $M=[M_{ij}]$ (with $i, j$ = 1, ..., $d$), is a $d\times d$ array from the numbers $ 1,...,d $ such that each number occurs $d$ times.
A Latin square of order $d$, denoted by $M=[M_{ij}]$ (with $i, j$ = 1, ..., $d$), is a $d\times d$ array from the numbers $ 1,...,d $ such that each number occurs in every row and every column only once. A {\bf row-Latin square} of order $d$ is a square, where each row is a permutation of the $d$ numbers \cite{Mullen-1998}. A {\bf column-Latin square} of order $d$ is a square, where each column is a permutation of the $d$ numbers. An example of a row-Latin square of order 4 is shown in Fig. \ref{patrat-boa}.

\begin{figure}[h]
\begin{center}
\begin{tabular}{|c|c|c|c|}
\hline   3 & 4 & 2 & 1 \\
\hline  3 & 4 & 1 & 2\\
\hline  2 & 1 & 4 & 3\\
\hline  1 & 3 & 2 & 4 \\
\hline
\end{tabular}
\end{center}
\caption{A row-Latin square of order 4.}
\label{patrat-boa}
\end{figure}

Two squares $M$ and $N$ are called orthogonal if all the pairs $(M_{ij}, N_{ij})$ are distinct.

In this paper, we analyze the special case when $d$ is a power of a prime number. An equivalent definition of a square is to assign to each number in the square an element $(x_1,x_2)$ in the space $\mathbb{F}_d\times \mathbb{F}_d$, $\mathbb{F}_d$ being the finite field with $d$ elements. By $x_1$ we label the row and by $x_2$ the column of the square.

{\bf Definition 2.1.} {\bf A square of order $d$} (with $d$ a power of a prime number $d=p^n$) is a partition of $\mathbb{F}_d\times \mathbb{F}_d$, being denoted as $S=\{ A_1, A_2,..., A_d\}$, where the cardinality of  $A_j$ is $d$.
To the elements of the subset $A_j$ in the square we assign the number $j$:
\beqa
&&\mbox{the}\; \mbox{elements}\; \mbox{of} \; A_1 \longrightarrow 1\nonumber \\
&&\mbox{the}\; \mbox{elements}\; \mbox{of} \; A_2 \longrightarrow 2\label{nr-asociat} \\
&& ...\nonumber \\
&&\mbox{the}\; \mbox{elements}\; \mbox{of} \; A_d \longrightarrow d. \nonumber
\eeqa

We show in Fig. \ref{Lat-ex} the definition of a Latin square of order 4 as a partition, where we have denoted by $\mu $ a primitive element of $\mathbb{F}_4$. The elements of $\mathbb{F}_4$ are: $\{ 0, 1, \mu , \mu^2 \}.$

\begin{figure}[h]
%\begin{figure*}[!ht]
\begin{center}
\begin{tabular}{c||c|c|c|c||}
\hline \hline $ \mu^2$ & 4 & 3 & 2 & 1\\
\hline $\mu$ & 3 & 4 & 1 & 2\\
\hline 1 & 2 & 1 & 4 & 3\\
\hline 0 & 1 & 2 & 3 & 4 \\
\hline   \hline   &0&1&$\mu$&$\mu^2$\\
\end{tabular}
\end{center}
\caption{ A Latin square described as a partition of $\mathbb{F}_4\times \mathbb{F}_4$: $A_1=\{(0,0)$, $(1,1)$, $(\mu ,\mu )$, $(\mu^2,\mu^2)\}$; $A_2=\{(0,1)$, $(1,0)$, $(\mu ,\mu^2)$, $(\mu^2,\mu )\}$; $A_3=\{(0,\mu )$, $(1,\mu^2)$, $(\mu ,0)$, $(\mu^2,1)\}$; $A_4=\{(0,\mu^2)$, $(1,\mu )$, $(\mu ,1)$, $(\mu^2,0)\}$.}
\label{Lat-ex}
\end{figure}

The concept of a partition of a set is equivalent to that of an equivalence relation on that set. In the case of a square, the equivalence classes are $A_1, A_2,..., A_d$.

Consider two squares of order $d$: $S=\{A_1, A_2,..., A_d\}$ and $\tilde S=\{B_1$, $B_2$,..., $B_d\}$.
If $a\in \mathbb{F}_d\times \mathbb{F}_d$, then we denote by $N(a)$ the number  $\in \{1,...,d\}$ associated to the element $a$ in the square $S$, while $\tilde N(a)$ is the associated number in the second square $\tilde S$. According to the rule \eqref{nr-asociat} we have: if $a\in A_k$, then $N(a)=k$, while for $ a\in B_m$, we get $\tilde N(a)=m$.

{\bf Definition 2.2.} Two squares $S$ and $\tilde S$ are called {\bf orthogonal} if
\[
(N(v),\tilde N(v))=(N(v_0),\tilde N(v_0))\Longleftrightarrow v=v_0,
\]
$v$ and $v_0$ being two elements in $\mathbb{F}_d\times \mathbb{F}_d$.\\
This is equivalent to the condition that all the pairs are distinct.

We denote by $\hat v$ the equivalence class of $v$ in the square $S$. The condition that the same number is assigned to both $v$ and $v_0$, i.e. $N(v)=N(v_0)$, leads to the conclusion that the two elements belong to the same equivalence class $\hat v=\hat v_0$. Further we denote by $\hat {\tilde v}$ the equivalence class of $v$ in the second square $\tilde S$. Then, the Definition 2.2 is equivalent to:

Two squares $S$ and $\tilde S$ are called orthogonal if
\beq
\hat v=\hat v_0\; \; \mbox{and} \; \; \hat {\tilde v}={\hat {\tilde v}}_0\ \Longleftrightarrow v=v_0.
\label{patr-ortog}
\eeq

{\bf Remark 2.3.} Definition 2.1 can be also given for arbitrary $D$, which is not necessary a power of a prime number. Let $M$ be a set with $D$ elements. A square of order $D$ is a partition of $M\times M$, being denoted as $S=\{ A_1, A_2,..., A_D\}$. The cardinality of each $A_j$ is $D$.

Definition 2.2 is the same for arbitrary $D$, which is not necessary a power of a prime number.

\section{Supersquares}

\subsection{Definition of the supersquares}

{\bf Definition 3.1.} Consider a square of order $d$ denoted by $S=\{A_1$, $A_2$, ..., $A_d\}$. $S$ is called a {\bf supersquare of order d} if $A_1$ is a subgroup with $d$ elements of $\mathbb{F}_d\times \mathbb{F}_d$ and $S=\mathbb{F}_d\times \mathbb{F}_d/A_1$ is the quotient set. The subsets $A_j$ are as follows:
\beqa
&& A_1 = \hat{0}; \nonumber \\
&& A_2= a_2+A_1 = \hat a_2; \nonumber \\
&& .... \nonumber \\
&& A_d= a_d+A_1 = \hat a_d,\nonumber
\eeqa
where $a_j\in \mathbb{F}_d\times \mathbb{F}_d$, $j$= 2, 3, ..., $d$.  Therefore, $S$ is well determined if the subgroup $A_1$ is given. $A_1$ is called the generating subgroup of the supersquare.

The Latin square given in Fig. \ref{Lat-ex} is a supersquare. The generating subgroup is $A_1=\{(0,0),(1,1),(\mu ,\mu ),(\mu^2 ,\mu^2)\}$ and $A_2=(0,1)+A_1$, $A_3=(0,\mu )+A_1$, $A_4=(0,\mu^2)+A_1$ as one can see in Fig. \ref{constructie}.

\begin{figure*}[!ht]
\begin{center}
\begin{tabular}{|c|c|c|c|}
\hline   & & & {\bf 1} \\
\hline   &  & {\bf 1} & \\
\hline   & {\bf 1} & & \\
\hline  {\bf 1} &  &  &  \\
\hline
\end{tabular}
\hspace{1cm}
\begin{tabular}{|c|c|c|c|}
\hline   & & 2 &\\
\hline   &  &  & 2 \\
\hline  2 & &  & \\
\hline   & 2 &  &  \\
\hline
\end{tabular}
\hspace{1cm}
\begin{tabular}{|c|c|c|c|}
\hline   & 3 & & \\
\hline  3 &  & & \\
\hline  &  &  & 3 \\
\hline   &  & 3 &  \\
\hline
\end{tabular}
\hspace{1cm}
\begin{tabular}{|c|c|c|c|}
\hline  4 & & & \\
\hline  & 4 &  & \\
\hline   &  & 4 & \\
\hline   &  &  & 4 \\
\hline
\end{tabular}
\vspace{0.2cm} \\
a) \hspace{3.1cm} b) \hspace{3.1cm} c) \hspace{3.1cm} d)
\end{center}
\caption{The Latin supersquare of order 4 given in Fig. \ref{Lat-ex}: a) the generating subgroup $A_1=\{(0,0),(1,1),(\mu ,\mu ),(\mu^2 ,\mu^2)\}$, b) $A_2=(0,1)+A_1$, c) $A_3=(0,\mu )+A_1$, d) $A_4=(0,\mu^2)+A_1$. }
\label{constructie}
\end{figure*}

{\bf Remark 3.2.} Definition 3.1 can be given for arbitrary $D$ if instead of $\mathbb{F}_d$ one uses a commutative group $(M,+)$ with $D$ elements and is as follows. A square of order $D$ is called a supersquare if there exists a subgroup $A_1$ of $M\times M$ with $D$ elements and $S=M\times M/A_{1}$.

\subsection{The construction of the mutually orthogonal supersquares of order $d$}
\label{constr-mos}

{\bf Proposition 3.3} {\it Suppose that $A_1$ and $B_1$ are subgroups with $d$ elements of  $\mathbb{F}_d\times \mathbb{F}_d$. Let $S$ and $\tilde S$ be the supersquares of order $d$ which are generated by $A_1$ and $B_1$. The squares $S$ and $\tilde S$ are orthogonal if and only if $A_1\cap B_1=\{0\} $.}

{\bf Proof.} We denote by  " $\hat{}$ " the cosets in $S$ and by  " $\hat{{\tilde{}}}$ "  the cosets in $\tilde S $.

"$\Leftarrow $" We have that $A_1\cap B_1=\{0\} $. Let us start with the conditions $\hat v=\hat v_0$ and $\hat {\tilde v}={\hat {\tilde v}}_0$. We obtain $\widehat{v-v_0}=\hat 0$. Since we have that $\hat 0=A_1$, we obtain $v-v_0 \in A_1$. The second condition is equivalent to
$v-v_0 \in \hat{\tilde{0}}=B_1$. Therefore we obtain $v-v_0\in A_1\cap B_1=\{0\}$, i.e. $v=v_0$. According to Eq. \eqref{patr-ortog}, the two squares are orthogonal.

"$\Rightarrow $" We know that the two squares are orthogonal. Suppose now that the intersection $A_1\cap B_1\ne \{0\} $; it means that there exists $z\ne 0$ with $z\in A_1\cap B_1$:
$$
z\in A_1=\hat 0\; \mbox{or} \; \mbox{equivalent} \; \hat z=\hat 0; \; \mbox{and} \;
z\in B_1=\hat{\tilde{0}} \; \mbox{or} \; \mbox{equivalent} \; \hat{\tilde{z}}=\hat{\tilde{0}}.
$$
Since the two squares are orthogonal, according to \eqref{patr-ortog}, we obtain that $z=0$, which is not possible since we start with $z\ne 0$.  $\Box $

{\bf Remark 3.4.} Proposition 3.3 is also valid for arbitrary $D$ with the replacement M instead of $\mathbb{F}_d$ (see Remark 3.2). The proof remains unchanged.

{\bf Proposition 3.5.} {\it (a) Consider $\{ v_1,v_2\}$ a basis in $\mathbb{F}_d\times \mathbb{F}_d$. Let us define the following subgroups with $d$ elements:
\beqa
A_1^{(j)}&:=&\mathbb{F}_d(v_1+\lambda_j\, v_2), \hspace{0.2cm} j=0,1,...,d-1; \nonumber \\
A_1^{(d)}&:=&\mathbb{F}_dv_2, \label{A-1}
\label{auri}
\eeqa
where $\lambda_j\in \mathbb{F}_d$ are all distinct. Then, the supersquares which are generated by the $d+1$ subgroups from \eqref{A-1} are mutually orthogonal.

(b) The maximum number of mutually orthogonal supersquares of order $d$ is $d+1$. The set of $d+1$ mutually orthogonal supersquares of order $d$ is called complete.}

{\bf Proof.} (a) One can easily prove that the intersection of any two of the subgroups \eqref{A-1} is the element zero. Therefore the set of $d+1$ supersquares defined by the generating subgroups $A_1^{(j)}$, with $j$ = 0, 1,..., $d$, are mutually orthogonal according to Proposition 3.3.

(b) According to Proposition 3.3, in order to obtain $m$ mutually orthogonal supersquares we need to construct their $m$ generating subgroups, with the property that the only common element of any two such subgroups is zero. Each generating subgroup has $d-1$ nonzero elements. Therefore, the total number of nonzero elements of the $m$ generating subgroups is $(d-1)\, m$. Since the total number of nonzero elements of $\mathbb{F}_d\times \mathbb{F}_d$ is $d^2-1$, we get: $(d-1)\, m\le d^2-1$, which leads to $m\le d+1$. The existence of the $d+1$ mutually orthogonal supersquares is shown in (a). $\Box $

{\bf Corollary 3.6.} $\mathbb{F}_d\times \mathbb{F}_d$ can be written as the union of $d+1$ subgroups with $d$ elements such that the intersection of any two of these subgroups is the element zero:
\[
\mathbb{F}_d\times \mathbb{F}_d=\bigcup_{j=0}^dA_1^{(j)}.
\]
Note that here $A_1^{(j)}$ are not necessary the subgroups of Eq. \eqref{A-1}.

According to Proposition 3.3 and Corollary 3.6, in order to obtain the complete set of $d+1$ mutually orthogonal supersquares, it is sufficient to obtain the $d+1$ generating subgroups $A_1^{(1)}$, $A_1^{(2)}$,..., $A_1^{(d)}$ such that the intersection of any two of these subgroups is the element zero. The $d+1$ mutually orthogonal supersquares are: $\mathbb{F}_d\times \mathbb{F}_d/A_1^{(1)}$, $\mathbb{F}_d\times \mathbb{F}_d/A_1^{(2)}$,..., $\mathbb{F}_d\times \mathbb{F}_d/A_1^{(d)}$.

{\bf Example.}

In the case $d=4$, let $A_1$, $B_1$, $C_1$, $D_1$, and $E_1$ be the generating subgroups of the five mutually orthogonal supersquares such that $\mathbb{F}_4\times \mathbb{F}_4 = A_1\cup B_1 \cup C_1 \cup D_1 \cup E_1 $ with $ X\cap Y =\{0\} $ for all $X$ and $Y$ two sets from $A_1$, $B_1$, $C_1$, $D_1$, $E_1$ ($X\ne Y$). Let $A_1 = \{ 0, a_1, a_2, a_3\}$ and $B_1 = \{ 0, b_1, b_2, b_3\}$, where $a_k\ne b_j$ for all $k$, $j$ = 1, 2, 3. The five mutually orthogonal supersquares are the following: $\mathbb{F}_d\times \mathbb{F}_d/A_1=\{A_1,A_1+b_1,A_1+b_2,A_1+b_3\}$, $\mathbb{F}_d\times \mathbb{F}_d/B_1=\{B_1,B_1+a_1,B_1+a_2,B_1+a_3\}$, $\mathbb{F}_d\times \mathbb{F}_d/C_1=\{C_1,C_1+a_1,C_1+a_2,C_1+a_3\}$, $\mathbb{F}_d\times \mathbb{F}_d/D_1=\{D_1,D_1+a_1,D_1+a_2,D_1+a_3\}$, $\mathbb{F}_d\times \mathbb{F}_d/E_1=\{E_1,E_1+a_1,E_1+a_2,E_1+a_3\}$.

In order to construct the generating subgroups, one has to start with $v_1, v_2\in \mathbb{F}_4\times \mathbb{F}_4$ such that $\{ v_1, v_2\}$ is a basis. If $\mu $ is a primitive element of $\mathbb{F}_4$, then one possible way of obtaining the generating subgroups is as follows: $A_1= \mathbb{Z}_2v_1+\mathbb{Z}_2v_2$, $B_1=\mu \, A_1$, $C_1=\mu^2 \, A_1$, $D_1=\mathbb{F}_4(v_1+\mu \, v_2)$, and $E_1=\mathbb{F}_4(v_1+\mu^2 \, v_2)$.
As an example we take $v_1=(1,0)$ and $v_2=(0,1)$ and obtain the complete set of mutually orthogonal supersquares shown in Fig. \ref{ex-ortog}.

\begin{figure*}[!ht]
\begin{center}
\begin{tabular}{|c|c|c|c|}
\hline 3 & 3 & 4 & 4 \\
\hline  3 &  3 & 4 & 4 \\
\hline  {\bf 1} & {\bf 1} & 2 & 2  \\
\hline   {\bf 1} & {\bf 1} & 2 & 2  \\
\hline
\end{tabular}
\hspace{0.2cm}
\begin{tabular}{|c|c|c|c|}
\hline 3 & 4 & 3 & 4 \\
\hline  {\bf 1} & 2 & {\bf 1} & 2 \\
\hline  3 & 4 & 3 & 4  \\
\hline  {\bf 1} & 2 & {\bf 1} & 2 \\
\hline
\end{tabular}
\hspace{0.2cm}
\begin{tabular}{|c|c|c|c|}
\hline  {\bf 1} & 2 & 2 & {\bf 1} \\
\hline  3 & 4 & 4 & 3 \\
\hline  3 & 4 & 4 & 3  \\
\hline  {\bf 1} & 2 & 2 & {\bf 1}  \\
\hline
\end{tabular}
\hspace{0.2cm}
\begin{tabular}{|c|c|c|c|}
\hline 4 & 3 & {\bf 1} & 2 \\
\hline  2 & {\bf 1} & 3 & 4 \\
\hline  3 & 4 & 2 & {\bf 1}  \\
\hline  {\bf 1} & 2 & 4 & 3  \\
\hline
\end{tabular}
\hspace{0.2cm}
\begin{tabular}{|c|c|c|c|}
\hline 2 & {\bf 1} & 4 & 3 \\
\hline  4 & 3 & 2 & {\bf 1} \\
\hline  3 & 4 & {\bf 1} & 2  \\
\hline  {\bf 1} & 2 & 3 & 4  \\
\hline
\end{tabular}
\vspace{0.2cm}\\
a) \hspace{2.4cm} b) \hspace{2.4cm} c) \hspace{2.4cm} d) \hspace{2.4cm} e)
\end{center}
\caption{A complete set of mutually orthogonal supersquares of order 4. The generating subgroups are denoted by bold 1. The supersquares d) and e) are Latin.}
\label{ex-ortog}
\end{figure*}

\section{Extraordinary supersquares}

In this section we introduce the concept of extraordinary supersquare of order $d$. This plays an important role in the field of quantum information theory, for example in the construction of mutually unbiased bases.

\subsection{Definition of the extraordinary supersquares. Classification of the squares of order $d$}

Let us consider $v_1=(x_1,y_1)$ and $v_2=(x_2,y_2)$ $\in \mathbb{F}_d\times \mathbb{F}_d$ with $d = p^n$. We denote by $\vert v_1\hspace{0.3cm} v_2 \vert$ the following determinant:
$$ \vert v_1\hspace{0.3cm} v_2 \vert = \left\vert
\begin{array}{cc}
x_1 & x_2\\
y_1 & y_2
\end{array}
\right\vert . $$

For $\alpha \in \mathbb{F}_{p^n}$, the trace is given by $ \mbox{tr} \; \alpha = \alpha +\alpha ^p + \alpha ^{p^2}+ ... + \alpha^{p^{n-1}}$.

We denote by $K$ the subgroup of $\mathbb{F}_{p^n}$, whose elements have the trace equal to zero:
\beq
K=\{ \alpha \in \mathbb{F}_{p^n} : \hspace{0.1cm} \mbox{tr} \; \alpha =0\}.
\label{k}
\eeq

{\bf Definition 4.1.} The subgroup $G\in \mathbb{F}_{p^n}\times \mathbb{F} _{p^n}$ is called {\bf extraordinary} if for any two of its elements $g_1$ and $g_2 \in G$, one has $\vert g_1 \hspace{0.3cm} g_2\vert \in K$.

The importance of the extraordinary subgroup is explained briefly in the Introduction, where we emphasized that the condition $tr(x_1y_2-x_2y_1)=0$ is equivalent to the commutation of two operators, whose associated elements are  $(x_1,y_1)$ and $(x_2,y_2)$ in $\mathbb{F}_d\times \mathbb{F}_d$ \cite{Klimov-2007}, \cite{Ghiu-2012}.

{\bf Definition 4.2.} A square of order $d$, $S=\{A_1,...,A_d\}$ is called {\bf extraordinary} if there is $j\in \{1,2,...,d\}$ such that $A_j$ is an extraordinary subgroup of $\mathbb{F}_d\times \mathbb{F}_d$.

As a consequence, a supersquare is extraordinary if its generating subgroup is extraordinary. The classification of the squares of order $d$ is shown in Table \ref{clasificare}.

\begin{table*}
\begin{center}
\begin{tabular}{|c|c|c|}
\hline \hline  Type of square & Description \\
\hline \hline
 Square  & $S=A_1\cup A_2...\cup A_d$\; with $A_j$ mutually \\
& disjoint subsets, each of cardinality $d$. \\
\hline   Supersquare & $A_1$ is a subgroup and $A_j= a_j+A_1$ \\
& with $j=2, 3,..., d$.\\
\hline   Extraordinary squares & $A_1$ is an extraordinary subgroup. \\
\hline   Extraordinary supersquares & $A_1$ is an extraordinary subgroup \\
 & and $A_j= a_j+A_1$ with $j=2, 3,..., d$. \\
\hline \hline
\end{tabular}
\end{center}
\caption{Classification of the squares of order $d$.}
\label{clasificare}
\end{table*}

{\bf Proposition 4.3.} Let $p$ be a prime number. Then any subgroup with $p$ elements of $\mathbb{F}_p\times \mathbb{F}_p$ is extraordinary.

{\bf Proof.} Let $G$ be a subgroup of $\mathbb{F}_p\times \mathbb{F}_p$, whose order is equal to $p$. Suppose $g\in G$ with $g\ne 0$. Then $ord(g) =p$. We know that $ord(g)$ divides the order of $G$: $ord(g)|p$. This leads to $\{ 0,1 g, 2 g,..., (p-1)g\}\subseteq G$. This means that $G=\mathbb{F}_pg$, which is an extraordinary subgroup. $\Box $

In the case of $p$ prime, according to Proposition 4.3, one can easily be proven that the complete set of $p+1$ mutually extraordinary supersquares have the form of Eq. \eqref{auri}, where $\{ v_1,v_2\}$ is a basis in $\mathbb{F}_p\times \mathbb{F}_p$.

\subsection{The construction of all the complete sets of mutually orthogonal extraordinary supersquares of order 4}

In this subsection we investigate the case $d=4$. The case $d$ being a prime number was discussed in Proposition 4.3, where we proved that the set of $d+1$ generating extraordinary subgroups have only the trivial form of Eq. \eqref{auri}. Case $d=4$ is the smallest order where the extraordinary subgroups cannot be written in the form $ \mathbb{F}_du$ (see Proposition 4.7 below). In the quantum information theory, the case $d=4$ corresponds to two-spin 1/2 systems. For higher order case, e.g. $d=2^3$, that represents systems with three particles of spin-1/2, we found four types of complete sets of mutually orthogonal extraordinary supersquares. The results for $d=8$ or higher will be analyzed elsewhere. The extraordinary subgroups play a central role for obtaining the mutually unbiased bases, which are important tools for quantum tomography, quantum cryptography, or quantum error correction codes.

{\bf Lemma 4.4.} {\it Consider $v_1\in \mathbb{F}_{2^n}\times \mathbb{F}_{2^n}$ with $v_1\ne 0$. Let $v_2\in \mathbb{F}_{2^n}\times \mathbb{F}_{2^n}$ such that $\{ v_1,v_2\} $ is a basis in $\mathbb{F}_{2^n}\times \mathbb{F}_{2^n} $. With the notation $\delta :=\vert v_1\hspace{0.3cm} v_2 \vert $, we have the following:

(a) if $x\in \mathbb{F}_{2^n}$, then
$\{w\in \mathbb{F}_{2^n}\times \mathbb{F}_{2^n} : \hspace{0.1cm} \vert v_1\hspace{0.3cm} w \vert =x\} = \delta^{-1}\, x\, v_2+\mathbb{F}_{2^n}\, v_1. $

(b) $\{w\in \mathbb{F}_{2^n}\times \mathbb{F}_{2^n} : \hspace{0.1cm} \vert v_1\hspace{0.3cm} w \vert \in K \} = K \delta^{-1}\,  v_2+\mathbb{F}_{2^n}\, v_1 $, where $K$ is given by \eqref{k} for $p=2$.

(c) If $x\in \mathbb{F}_{2^n}$, then $\exists $ a unique element $w\in \mathbb{F}_{2^n} v_2$ such that $\vert v_1\hspace{0.3cm} w \vert =x$. This element is $w=\delta^{-1}x\, v_2$.}

{\bf Corollary 4.5.} Consider a nonzero element $v_1\in \mathbb{F}_{2^n}\times \mathbb{F}_{2^n}$ and $x\in \mathbb{F}_{2^n}$ with $x\ne 0$. Then there is $\tilde v_2\in \mathbb{F}_{2^n}\times \mathbb{F}_{2^n}$ such that $\vert v_1\hspace{0.3cm} \tilde v_2 \vert =x$.

{\bf Lemma 4.6.} {\it Let $G \subseteq \mathbb{F}_{2^n}\times \mathbb{F}_{2^n}$ be a subgroup (with $G\ne 0$). Then we have the following:

(a) If $\exists \; v\in \mathbb{F}_{2^n}\times \mathbb{F}_{2^n} $, with $v\ne 0$ such that $G\subseteq \mathbb{F}_{2^n}v$, then $G$ is an extraordinary subgroup.

(b) Suppose that $G\ne 0$ is an extraordinary subgroup such that $G\nsubseteq \mathbb{F}_{2^n}v  $, $\forall \; v\in \mathbb{F}_{2^n}\times \mathbb{F}_{2^n}$. Let $v_1\in G\, \backslash \, \{0\}$ and $v_2\in G\, \backslash \, \mathbb{F}_{2^n}v_1$. With the notation $\delta :=\vert v_1\hspace{0.3cm} v_2 \vert $, we have:

\hspace{0.5cm} (i) $\delta \in K\, \backslash \, \{0\}$;

\hspace{0.5cm} (ii) $G\subseteq K\, \delta^{-1}\, v_1+K\, \delta^{-1}\, v_2$, where $K$ is given by \eqref{k} for $p=2$.}

Proposition 4.7 below gives the most general expression of an extraordinary subgroup in $\mathbb{F}_4\times \mathbb{F}_4$.

{\bf Proposition 4.7.} {\it Suppose that $G\subseteq \mathbb{F}_4\times \mathbb{F}_4$ is a subgroup which contains four elements. Let $v$ be a nonzero element of $G$. Consider $w\in \mathbb{F}_4\times \mathbb{F}_4$ such that $\vert v\hspace{0.3cm} w \vert = 1$ (there is such $w$ according to Corollary 4.5).}
{\it Then we have that $G$ is extraordinary if and only if:}
\beqa
&& \mbox{(i)}\; G = \mathbb{F}_4 \, v \hspace{0.3cm} or
\nonumber \\
&& \mbox{(ii)}\; G = \mathbb{Z}_2\, v + \mathbb{Z}_2\, (w+\lambda v), \hspace{0.3cm} with \; \lambda \in \{0,\mu \},
\nonumber
\eeqa
{\it where $\mu $ is the primitive element of $\mathbb{F}_4$.}

{\bf Proof.} The proof of Proposition 4.7 is given in Appendix A.

{\bf Lemma 4.8.} {\it  Suppose that $A_1, B_1, C_1, D_1,$ and $E_1$ are extraordinary subgroups, which contain four elements, such that $ \mathbb{F}_4\times \mathbb{F}_4 = A_1\cup B_1 \cup C_1 \cup D_1 \cup E_1 $ and $ X\cap Y =\{0\} $ for all $X$ and $Y$ two sets from $A_1$, $B_1$, $C_1$, $D_1$, $E_1$ with $X\ne Y$. Then, at least one of the five subgroups has the form $\mathbb{F}_4\, u$, with $u\in \mathbb{F}_4\times \mathbb{F}_4$, $u\ne 0$.}

{\bf Proof.} The proof of Lemma 4.8 is given in Appendix B.

According to Proposition 3.3 and Corollary 3.6, we have to write the whole space $ \mathbb{F}_4\times \mathbb{F}_4$ as a union of five extraordinary subgroups, such that the intersection of any two of them is the element zero. Further we apply Lemma 4.8 in the following two theorems, i.e. we know that at least one of the five subgroups has the form $\mathbb{F}_4\, u$.

 Theorem 4.9 below says that if two of the extraordinary subgroups which construct the space $ \mathbb{F}_4\times \mathbb{F}_4$ are of the form $\mathbb{F}_4\, u$, then also the other three subgroups must have the same structure $\mathbb{F}_4\, w$.

{\bf Theorem 4.9.} {\it Suppose that $A_1, B_1, C_1, D_1,$ and $E_1$ are extraordinary subgroups, which contain four elements, such that $ \mathbb{F}_4\times \mathbb{F}_4 = A_1\cup B_1 \cup C_1 \cup D_1 \cup E_1 $ and $ X\cap Y =\{0\} $ for all $X$ and $Y$ two sets from $A_1$, $B_1$, $C_1$, $D_1$, $E_1$ with $X\ne Y$. If there exists $v_1$ and $v_2\in \mathbb{F}_4\times \mathbb{F}_4 $ such that  $A_1= \mathbb{F}_4 \, v_1$ and $B_1=\mathbb{F}_4\, v_2$, then $C_1$, $D_1$, $E_1$ are the following}
\beq
 \mathbb{F}_4\, (v_1+ \mu \, v_2);\hspace{0.2cm}
\mathbb{F}_4\, (v_1+ \mu ^2\, v_2); \hspace{0.2cm}
\mathbb{F}_4\, (v_1+ v_2)
\label{prop-5}
\eeq
{\it or permuted. $\mu $ is a primitive element of $\mathbb{F}_4$.}

{\bf Proof.} Since $A_1\cap B_1=\{0\}$, we obtain that $\{v_1,v_2\}$ is a basis in $ \mathbb{F}_4\times \mathbb{F}_4$. We denote the nonzero parameter $\delta :=\vert v_1\hspace{0.3cm} v_2 \vert$. Then, by using Lemma 4.4 (c), there exists $u$ given by $u = \delta ^{-1}\, v_2$, such that $\vert v_1\hspace{0.3cm} u \vert = 1$. It is obvious that $u\in B_1$ and that the subgroup $B_1$ can equivalently be written as $B_1 = \mathbb{F}_4 u$.

Because $\{v_1,v_2\}$ is a basis, one obtains that
$v_1 + \mu \, u \notin A_1 \cup B_1$.
This means that $v_1 + \mu \, u$ belongs to one of the subgroups
$C_1$, $D_1$, $E_1$; let us take $v_1 + \mu \, u \in C_1$.
We have $\vert v_1 + \mu \, u \hspace{0.3cm} u \vert = 1.$
We apply Proposition 4.7, where $v$ is replaced by $v_1 + \mu \, u$ and $w$ by $u$.
Therefore $C_1=\mathbb{F}_4\, (v_1+ \mu \, u)$ according to (i) or
\beq
\label{12C1}
C_1 = \mathbb{Z}_2\, (v_1+\mu \, u) + \mathbb{Z}_2\, [u+\lambda (v_1+\mu \, u)],
\eeq
according to (ii), where $\lambda \in \{0, \mu \}$.

Let us assume that equality \eqref{12C1} is true.
If $\lambda =0$, then $u\in C_1$, which is false since $u\in B_1$ and $u \ne 0$
($C_1 \cap B_1=\{0\}$).
If $\lambda =\mu $, then
$
C_1 = \mathbb{Z}_2\, (v_1+\mu \, u) + \mathbb{Z}_2\, (\mu \, u+\mu \, v_1)
$ $\Rightarrow$ $(v_1+\mu \, u)+(\mu \, u+\mu \, v_1)=\mu^2 v_1$$\in C_1$,
which is false since $\mu^2 v_1 \in A_1$ and $\mu^2 v_1 \ne 0$
($C_1 \cap A_1=\{0\}$). This means that the equality \eqref{12C1} is false. Therefore we have
$C_1=\mathbb{F}_4\, (v_1+ \mu \, u)$.

Because $\{v_1,v_2\}$ is a basis, one easily notice that
$v_1 + \mu ^2 u \notin A_1 \cup B_1 \cup C_1$.
This means that $v_1 + \mu ^2  u$  belongs to one of the subgroups $D_1$, $E_1$; let us take $v_1 + \mu ^2 u \in D_1$.
We have $\vert v_1 + \mu ^2 u \hspace{0.3cm} u \vert = 1.$
We apply Proposition 4.7 and we obtain $D_1=\mathbb{F}_4\, (v_1+ \mu ^2 u )$ (analogously as above).

Now, we prove that
\beq
v_1+u, \mu (v_1+u), \mu^2 (v_1+u) \notin A_1\cup B_1\cup C_1\cup D_1.
\label{ec-intermed}
\eeq

Suppose that there is $\beta \in \{1, \mu ,\mu^2\}$ and $X$ a subgroup from $A_1$, $B_1$, $C_1$, or $D_1$ such that $\beta (v_1+u) \in  X$. This leads to the fact that $v_1+u \in \beta^{-1}\, X = X$. We already proved that $v_1 + \alpha \, u \notin A_1, B_1$ for any $\alpha \in \{ 1,\mu , \mu^2\}$. We need to prove that $v_1+u \notin C_1 \cup D_1$.
Suppose that there is $\nu \in \mathbb{F}_4$ such that
$v_1 + u = \nu (v_1 +\mu \, u). $
This is equivalent to $ \nu = 1$ and $\mu \, \nu =1$, which cannot be simultaneously fulfilled. Therefore, we obtain that $v_1+u \notin C_1$. Suppose now that there is $\nu \in \mathbb{F}_4$ such that
$v_1 + u = \nu (v_1 +\mu^2 \, u)$.
This is equivalent to $ \nu = 1$ and $\mu ^2\, \nu =1$, which cannot be both verified. Therefore, we obtain that $v_1+u \notin D_1$. We proved (\ref{ec-intermed}).
Therefore $v_1+u$, $\mu (v_1+u)$, $\mu^2 (v_1+u)$ $\in E_1$, it means
$\mathbb{F}_4\, (v_1+ u)=E_1$.

We proved that the subgroups $C_1$, $D_1$, $E_1$ are
\begin{equation}
\label{subgrupurile}
\mathbb{F}_4\, (v_1+ \mu \, u),
\quad
\mathbb{F}_4\, (v_1+ \mu^2 u),
\quad
\mathbb{F}_4\, (v_1+ u),
\end{equation}
or permuted.
We have $u \in \{v_2, \, \mu \, v_2, \, \mu^2 v_2 \}$ and for these three values, the subgroups
 \eqref{subgrupurile} become
\beqa
&&\mbox{for  }
u=v_2:
\mathbb{F}_4\, (v_1+ \mu \, v_2),
\quad
\mathbb{F}_4\, (v_1+ \mu^2 v_2),
\quad
\mathbb{F}_4\, (v_1+ v_2),\nonumber \\
&&
\mbox{for  }
u=\mu \, v_2:
\mathbb{F}_4\, (v_1+ \mu^2 v_2),
\quad
\mathbb{F}_4\, (v_1+ v_2),
\quad
\mathbb{F}_4\, (v_1+ \mu \, v_2),\nonumber \\
&&
\mbox{for  }
u=\mu^2 v_2:
\mathbb{F}_4\, (v_1+ v_2),
\quad
\mathbb{F}_4\, (v_1+ \mu \, v_2),
\quad
\mathbb{F}_4\, (v_1+ \mu^2 v_2),\nonumber
\eeqa
i.e. the subgroups of Eq. \eqref{prop-5}, or permuted. $\Box $

{\bf Remark 4.10.} The converse of Theorem 4.9 holds. If $\{v_1, v_2\}$ is a basis in $\mathbb{F}_4\times \mathbb{F}_4 $,
then the union of the subgroups $\mathbb{F}_4 \, v_1$, $\mathbb{F}_4\, v_2$ and the three subgroups \eqref{prop-5} is
$\mathbb{F}_4\times \mathbb{F}_4$ and the intersection of any two of them is zero (according to Proposition 3.5). These five subgroups are extraordinary (Lemma 4.6, (a)).

{\bf Theorem 4.11.} {\it Suppose that $A_1, B_1, C_1, D_1,$ and $E_1$ are extraordinary subgroups, which contain four elements, such that $ \mathbb{F}_4\times \mathbb{F}_4 = A_1\cup B_1 \cup C_1 \cup D_1 \cup E_1 $ and $ X\cap Y =\{0\} $ for all $X$ and $Y$ two sets from $A_1$, $B_1$, $C_1$, $D_1$, $E_1$ with $X\ne Y$. By using Lemma 4.8 we know that at least one of the five subgroups has the form $\mathbb{F}_4\, v_1$. Let us denote $A_1 = \mathbb{F}_4\, v_1$ and assume that $X \ne \mathbb{F}_4\, w$ for $X$ from $B_1$, $C_1$, $D_1$, $E_1$  for any $w\in \mathbb{F}_4 \times \mathbb{F}_4$. Consider $v_2 \in \mathbb{F}_4 \times \mathbb{F}_4$ such that $ \vert v_1\hspace{0.3cm} v_2 \vert = 1$ (there is such $v_2$ according to Corollary 4.5). Then, the other four extraordinary subgroups have the form:}
$$\mathbb{Z}_2\, v_2 + \mathbb{Z}_2\, (v_1+\mu \, v_2); \hspace{0.14cm}
\mathbb{Z}_2\, \mu \, v_2 + \mathbb{Z}_2\, (\mu^2\, v_1+\mu^2 \, v_2); $$
$$\mathbb{Z}_2\, \mu^2\, v_2 + \mathbb{Z}_2\, (\mu \, v_1+\mu \, v_2); \hspace{0.14cm}
\mathbb{Z}_2\, (v_1+v_2) + \mathbb{Z}_2\, (\mu \, v_1+\mu^2 \, v_2).$$

{\bf Proof.} The condition $\vert v_1\hspace{0.3cm} v_2 \vert = 1$ leads to $v_2\notin \mathbb{F}_4 v_1 = A_1$. Let us consider that $v_2\in B_1$. We know that the extraordinary subgroup $B_1\ne \mathbb{F}_4\, v_2$, therefore according to Proposition 4.7, $B_1$ is of type (ii):
$
B_1 = \mathbb{Z}_2\, v_2 + \mathbb{Z}_2\, (v_1+\lambda v_2),
$
where $\lambda \in \{ 0, \mu \}$. If $\lambda = 0$, then $ B_1 = \mathbb{Z}_2\, v_2 + \mathbb{Z}_2\, v_1$, which cannot be fulfilled because $v_1\in A_1$. This leads to the fact that $\lambda = \mu$ and further
$
B_1 = \mathbb{Z}_2\, v_2 + \mathbb{Z}_2\, (v_1+\mu v_2).
$
Since $\mu \, v_2\notin A_1\cup B_1$, we may consider that $\mu \, v_2\in C_1$. We have
$ \vert \mu \, v_2\hspace{0.3cm} \mu^2\, v_1 \vert = 1$.
From Proposition 4.7 we know that the extraordinary subgroup $C_1$ has the expression
$
C_1 = \mathbb{Z}_2\, \mu \, v_2 + \mathbb{Z}_2\, (\mu^2\, v_1+\mu^2 v_2),
$
where we used that $C_1\ne \mathbb{F}_4\, w$ and that $\lambda \ne 0$ (the same argumentation as above in the case $B_1$).

We have that $\mu^2 \, v_2\notin A_1\cup B_1\cup C_1$ and we take $\mu^2 \, v_2\in D_1 $. We get
$ \vert \mu^2 \, v_2\hspace{0.3cm} \mu \, v_1 \vert = 1$.
We apply Proposition 4.7 with $v=\mu^2 \, v_2$ and $w=\mu \, v_1$ and obtain
$$
D_1= \mathbb{Z}_2\, \mu^2\, v_2 + \mathbb{Z}_2\, (\mu \, v_1+ v_2)
=\{ 0, \mu^2\, v_2, v_2+\mu \, v_1, \mu (v_1+ v_2)\}
=\mathbb{Z}_2\, \mu^2\, v_2 + \mathbb{Z}_2\, (\mu \, v_1+\mu \, v_2).
$$
The case $\lambda = 0$ in Proposition 4.7, case (ii) is not allowed, as we already discussed above for $B_1$: the same idea.
There are three nonzero elements which are not included in $A_1\cup B_1\cup C_1\cup D_1$. We will denote by $E_1$ this subgroup:
$E_1=\mathbb{Z}_2\, (v_1+v_2) + \mathbb{Z}_2\, (\mu \, v_1+\mu^2 \, v_2)$.  $\Box $

{\bf Remark 4.12.} The converse of Theorem 4.11 holds. If $ \vert v_1\hspace{0.3cm} v_2 \vert = 1$ then the union of the five subgroups of Theorem 4.11 is $\mathbb{F}_4\times \mathbb{F}_4$ and the intersection of any two of them is zero.

{\bf Theorem 4.13.}
{\it We have an equality of the following kind:

$\mathbb{F}_4 \times \mathbb{F}_4 = A_1 \cup B_1 \cup C_1 \cup D_1 \cup E_1 $,
with $A_1$, $B_1$, $C_1$, $D_1$, $E_1$ extraordinary subgroups,
and $ X \cap Y =\{ 0\} $ for all two different sets $X$, $Y$ from $A_1$, $B_1$, $C_1$, $D_1$, $E_1$,
if and only if one of the two cases below holds

I) $A_1= \mathbb{F}_4 \, v_1$; $B_1=\mathbb{F}_4\, v_2$; $C_1=\mathbb{F}_4\, (v_1+ \mu \, v_2)$;
$D_1=\mathbb{F}_4\, (v_1+ \mu ^2\, v_2)$;
$E_1=\mathbb{F}_4\, (v_1+ v_2)$, or permuted, where $\{ v_1,v_2\}$ is an arbitrary basis in $\mathbb{F}_4 \times \mathbb{F}_4$;

or

II) $A_1= \mathbb{F}_4 \, v_1$;
$B_1=\mathbb{Z}_2\, v_2 + \mathbb{Z}_2\, (v_1+\mu \, v_2) $; $C_1=\mathbb{Z}_2\, \mu \, v_2 + \mathbb{Z}_2\, (\mu^2\, v_1+\mu^2 \, v_2)$; $D_1=\mathbb{Z}_2\, \mu^2\, v_2 + \mathbb{Z}_2\, (\mu \, v_1+\mu \, v_2) $;
$E_1=\mathbb{Z}_2\, (v_1+v_2) + \mathbb{Z}_2\, (\mu \, v_1+\mu^2 \, v_2)$, or permuted, where $v_1$, $v_2\in \mathbb{F}_4 \times \mathbb{F}_4$ such that $ \vert v_1\hspace{0.3cm} v_2 \vert = 1$.}

{\bf Proof.} The condition that the whole space $ \mathbb{F}_4\times \mathbb{F}_4$ is written as a union of five extraordinary subgroups, such that the intersection of any two such subgroups is the element zero, leads to the conclusion that at least one of the five subgroups has the form $\mathbb{F}_4\, u$ due to Lemma 4.8. We will denote this extraordinary subgroup as $A_1= \mathbb{F}_4 \, v_1$, where $v_1\in \mathbb{F}_4 \times \mathbb{F}_4$.

If only one of the five extraordinary subgroups has the form $\mathbb{F}_4\, u$, then according to Theorem 4.11, the other four subgroups have the expression presented by Theorem 4.11, i.e. Type II). Otherwise, if there are two extraordinary subgroups of the form $\mathbb{F}_4\, u$, let denote them as $A_1= \mathbb{F}_4 \, v_1$ and $B_1= \mathbb{F}_4 \, v_2$, then the other three subgroups $C_1$, $D_1$, $E_1$ have the structure given by Theorem 4.9, i.e. we obtain Type I). $\Box $

\subsection{Examples of mutually orthogonal extraordinary supersquares of order 4}

We use Theorem 4.13 for giving two examples of complete set of mutually orthogonal extraordinary supersquares of order 4.

Type I: Let us consider $v_1=(1, \mu^2)$ and $v_2=(0,\mu )$. The complete set of mutually orthogonal extraordinary supersquares of order 4 which is generated is presented in Fig. \ref{ex-3-axe}. Type I mutually orthogonal extraordinary supersquares contain the complete set $d-1=3$ Latin orthogonal supersquares of order 4. In addition, there is the row-Latin square characterized by the same permutation in all rows and the column-Latin square characterized by the same permutation in all columns.

\begin{figure*}[!ht]
\begin{center}
\begin{tabular}{|c|c|c|c|}
\hline  3 & {\bf 1} & 2 & 4 \\
\hline  2 & 4 & 3 & {\bf 1} \\
\hline  4 & 2 & {\bf 1} & 3  \\
\hline  {\bf 1} & 3 & 4 & 2  \\
\hline
\end{tabular}
\hspace{0.2cm}
\begin{tabular}{|c|c|c|c|}
\hline  {\bf 1} & 2 & 3 & 4 \\
\hline  {\bf 1} & 2 & 3 & 4 \\
\hline  {\bf 1} & 2 & 3 & 4  \\
\hline  {\bf 1} & 2 &  3 & 4  \\
\hline
\end{tabular}
\hspace{0.2cm}
\begin{tabular}{|c|c|c|c|}
\hline 3 & 2 & 4 & {\bf 1} \\
\hline  2 & 3 & {\bf 1} & 4 \\
\hline  4 & {\bf 1} & 3 & 2  \\
\hline  {\bf 1} & 4 & 2 & 3  \\
\hline
\end{tabular}
\hspace{0.2cm}
\begin{tabular}{|c|c|c|c|}
\hline 2 & 2 & 2 & 2 \\
\hline   4 & 4 & 4 & 4 \\
\hline  3 & 3 & 3 & 3  \\
\hline  {\bf 1} & {\bf 1} & {\bf 1} & {\bf 1}  \\
\hline
\end{tabular}
\hspace{0.2cm}
\begin{tabular}{|c|c|c|c|}
\hline 4 & 2 & {\bf 1} & 3 \\
\hline  3 & {\bf 1} & 2 & 4 \\
\hline  2 & 4 & 3 & {\bf 1}  \\
\hline  {\bf 1} & 3 & 4 & 2  \\
\hline
\end{tabular}
\vspace{0.2cm}\\
a) \hspace{2.4cm} b) \hspace{2.4cm} c) \hspace{2.4cm} d) \hspace{2.4cm} e)
\end{center}
\caption{A complete set of mutually orthogonal extraordinary supersquares of order 4 of Type I. The basis which generates the supersquares is $v_1=(1, \mu^2)$ and $v_2=(0,\mu )$. The supersquares a), c) and e) are Latin.}
\label{ex-3-axe}
\end{figure*}

 For obtaining an example of Type II, we take $v_1=(1,\mu^2)$ and $v_2=(1,\mu )$. The complete set of mutually orthogonal extraordinary supersquares of order 4 generated is presented in Fig. \ref{ex-nu-axa}. The supersquare a), whose generating subgroup is $\mathbb{F}_4\, v_1$, is Latin.

\begin{figure*}[!ht]
\begin{center}
\begin{tabular}{|c|c|c|c|}
\hline  4 & {\bf 1} & 3 & 2 \\
\hline  3 & 2 & 4 & {\bf 1} \\
\hline  2 & 3 & {\bf 1} & 4  \\
\hline  {\bf 1} & 4 & 2 & 3  \\
\hline
\end{tabular}
\hspace{0.2cm}
\begin{tabular}{|c|c|c|c|}
\hline 3 & 2 & 2 & 3 \\
\hline  4 & {\bf 1} & {\bf 1} & 4 \\
\hline  2 & 3 & 3 & 2  \\
\hline  {\bf 1} & 4 & 4 & {\bf 1}  \\
\hline
\end{tabular}
\hspace{0.2cm}
\begin{tabular}{|c|c|c|c|}
\hline {\bf 1} & 2 & {\bf 1} & 2 \\
\hline  3 & 4 & 3 & 4 \\
\hline  3 & 4 & 3 & 4  \\
\hline  {\bf 1} & 2 & {\bf 1} & 2  \\
\hline
\end{tabular}
\hspace{0.2cm}
\begin{tabular}{|c|c|c|c|}
\hline 4 & 2 & 3 & {\bf 1} \\
\hline  {\bf 1} & 3 & 2 & 4 \\
\hline  4 & 2 & 3 & {\bf 1}  \\
\hline  {\bf 1} & 3 & 2 & 4  \\
\hline
\end{tabular}
\hspace{0.2cm}
\begin{tabular}{|c|c|c|c|}
\hline 2 & 2 & 4 & 4 \\
\hline  2 & 2 & 4 & 4 \\
\hline  {\bf 1} & {\bf 1} & 3 & 3  \\
\hline  {\bf 1} & {\bf 1} & 3 & 3  \\
\hline
\end{tabular}
\vspace{0.2cm}\\
a) \hspace{2.4cm} b) \hspace{2.4cm} c) \hspace{2.4cm} d) \hspace{2.4cm} e)
\end{center}
\caption{A complete set of mutually orthogonal extraordinary supersquares of order 4 of Type II. The elements which generate the supersquares are $v_1=(1,\mu ^2)$ and $v_2=(1,\mu )$. We have the following: a) Latin, b) column-Latin, c) general, d) row-Latin, and e) general. }
\label{ex-nu-axa}
\end{figure*}

\section{Conclusions}
In this paper we introduced the concept of supersquare of order $d$.  The main result presented here is the algorithm of construction of mutually orthogonal supersquares of order $d$ and the generation of all the complete sets of mutually orthogonal extraordinary supersquares of order 4.

\section*{Acknowledgements}
The work of Iulia Ghiu was supported by CNCS - UEFISCDI, postdoctoral research project PD code 151, no. 150/30.07.2010 for the University of Bucharest.

\appendix

\section{Proof of Proposition 4.7}

"$\Rightarrow $" For $d=2^2$ we obtain $K=\mathbb{Z}_2$, with $K$ given by \eqref{k}. There are two possible cases:

(i) $\exists \, u\in \mathbb{F}_4\times \mathbb{F}_4$ such that $G\subseteq  \mathbb{F}_4 \, u$ or

(ii) $\forall \, u\in \mathbb{F}_4\times \mathbb{F}_4 $ one has $G\nsubseteq \mathbb{F}_4 \, u$.

Case (i): Since $G$ and $\mathbb{F}_4 u$ have 4 elements, it means that $G=\mathbb{F}_4 u$. Since $v\in G \subseteq \mathbb{F}_4 \, u$, we obtain $v\in \mathbb{F}_4 \, u$; $\mathbb{F}_4 v\subseteq \mathbb{F}_4 u=G$. This leads to $G = \mathbb{F}_4 \, v$.

(ii) We have now that $\forall \, u\in \mathbb{F}_4\times \mathbb{F}_4 $ $\Rightarrow $ $G\nsubseteq \mathbb{F}_4 \, u$. This means that $G\nsubseteq \mathbb{F}_4 \, v$ $\Rightarrow $ $\exists \, \tilde v\in G\, \backslash \, \mathbb{F}_4 \, v$. From the condition $\vert v\hspace{0.3cm} w \vert = 1$, we obtain that $\{ v,w\}$ is a basis in $\mathbb{F}_4\times \mathbb{F}_4 $. Let us write $\tilde v$ as $\tilde v=\lambda v+\rho w$, where $\lambda ,\rho \in \mathbb{F}_4$.
Since $G$ is an extraordinary subgroup, one has $\vert v\hspace{0.3cm} \tilde v \vert \in \mathbb{Z}_2$;
$\tilde v \notin \mathbb{F}_4 \, v$$\Rightarrow$
$\{ v,\tilde v\}$ is a basis in $\mathbb{F}_4\times \mathbb{F}_4 $,
therefore $\vert v\hspace{0.3cm} \tilde v \vert \neq 0$;
one obtains $\vert v\hspace{0.3cm} \tilde v \vert =1$.
We compute $\vert v\hspace{0.3cm} \tilde v \vert=\lambda \vert v\hspace{0.3cm} v \vert+\rho \vert v\hspace{0.3cm} w \vert =\rho $. This means $\rho =1$, i.e. $\tilde v=w+ \lambda v$.

Further we apply Lemma 4.6 (b), where we replace $v_1$ by $v$ and $v_2$ by $\tilde v$ and obtain $G\subseteq \mathbb{Z}_2v+\mathbb{Z}_2\tilde v$. This is actually an equality, since the number of elements of both sets is four:
$G= \mathbb{Z}_2v+\mathbb{Z}_2(w+\lambda v)$, with $\lambda \in \mathbb{F}_4=\{ 0,1,\mu ,\mu^2 \}$. One can easily check that it is sufficient to choose $\lambda \in \{ 0,\mu \}$.

"$\Leftarrow $" (i) If $G = \mathbb{F}_4\, v$, then from Lemma 2, a) we obtain that $G$ is an extraordinary subgroup.

(ii) Suppose $G=\mathbb{Z}_2v+\mathbb{Z}_2(w+\lambda v)$. Let $g_1,g_2\in G$. Hence $\exists \, q_1,q_2\in \mathbb{Z}_2$ such that $g_1=q_1\, v+q_2\, (w+\lambda v)$ and $\exists \, q_3,q_4\in \mathbb{Z}_2$ such that $g_2=q_3\, v+q_4\, (w+\lambda v)$. Then $\vert g_1\hspace{0.3cm} g_2 \vert =q_1q_4+q_2q_3\in \mathbb{Z}_2=K$, i.e. $G$ is extraordinary. $\Box $

\section{Proof of Lemma 4.8}

Assume that none of the five subgroups $A_1$, $B_1$, $C_1$, $D_1,$ and $E_1$ has the expression $\mathbb{F}_4\, u$. Consider $v_1\in A_1$ such that $v_1\ne 0$. Let $w_1\in \mathbb{F}_4\times \mathbb{F}_4$ such that $\vert v_1 \hspace{0.3cm} w_1 \vert =1$. Then, according to Proposition 4.7, $\exists \, \lambda_1 \in \{0,\mu \}$ such that $A_1=\mathbb{Z}_2\, v_1 + \mathbb{Z}_2\, (w_1+\lambda_1 v_1)$. With the notation $v_2=w_1+\lambda_1 v_1$, we obtain $\vert v_1 \hspace{0.3cm} v_2 \vert =1$, which means that $\{ v_1,v_2\}$ is a basis in $\mathbb{F}_4\times \mathbb{F}_4$. We get $A_1=\mathbb{Z}_2\, v_1 + \mathbb{Z}_2\, v_2$.

Let us define now $v:=v_1+\mu\, v_2$. It is obvious that $v\notin A_1$. Suppose that $v\in B_1$. With the notation $w=\mu^2\, v_1$ we have $\vert v \hspace{0.3cm} w \vert =1$. By applying Proposition 4.7, we obtain $B_1=\mathbb{Z}_2\, v + \mathbb{Z}_2\, (w+\lambda v)=\mathbb{Z}_2\, (v_1+\mu v_2) + \mathbb{Z}_2\, [\mu^2v_1+\lambda (v_1+\mu v_2)]$, where $\lambda \in \{0,\mu \}$. Further we will prove that the case $\lambda =\mu $ is not possible. If $\lambda =\mu $, one obtains $v_1+\mu \, v_2+v_1+\mu^2 v_2=v_2\in B_1$. But we already know that $v_2\in A_1$, which leads to $v_2\in A_1\cap B_1$, which is not true. Therefore, we get $\lambda =0$ and the expression of $B_1$ is $B_1=\mathbb{Z}_2\, (v_1+\mu \, v_2) + \mathbb{Z}_2\, \mu^2 v_1$.

Let us define further $\tilde v:=v_2+\mu\, v_1$. It is obvious that $\tilde v\notin A_1\cup B_1$. Assume that $\tilde v\in C_1$. For $\tilde w=\mu^2 v_2$ we obtain $\vert \tilde v \hspace{0.3cm} \tilde w \vert =1$. With the help of Proposition 4.7 we write $C_1=\mathbb{Z}_2\, \tilde v + \mathbb{Z}_2\, (\tilde w+\tilde \lambda \, \tilde v)$, where $\tilde \lambda \in \{0,\mu \}$. We will prove that neither $\tilde \lambda =0$, neither $\tilde \lambda =\mu $ are allowed. If $\tilde \lambda =\mu $, one obtains $v_2+\mu \, v_1+v_2+\mu^2 v_1=v_1\in C_1$. But we already know that $v_1\in A_1$, which leads to $v_1\in A_1\cap C_1$, which is not true. The other possibility is $\tilde \lambda =0$, which leads to $C_1=\mathbb{Z}_2(v_2+\mu \, v_1)+\mathbb{Z}_2\mu^2 v_2$. One can readily check that $\mu \, v_1+\mu \, v_2 \in B_1\cap C_1$, which is not possible since the unique common element of $B_1$ and $C_1$ is zero. This leads to the fact that our assumption that none of the five subgroups $A_1$, $B_1$, $C_1$, $D_1,$ and $E_1$ has the expression $\mathbb{F}_4\, u$ is wrong.

\end{document}